\documentclass[twocolumn,showpacs,preprintnumbers,amsmath,amssymb]{revtex4}

\newcommand{\beq}{\begin{equation}}
\newcommand{\eeq}{\end{equation}}
\def\beqa{\begin{eqnarray}}
\def\eeqa{\end{eqnarray}}
\def\la{\lower.5ex\hbox{$\; \buildrel < \over \sim \;$}}
\def\ga{\lower.5ex\hbox{$\; \buildrel > \over \sim \;$}}
\def\fixit#1{}
\def\hmpc{h^{-1}\,{\rm Mpc}}
\def\hkpc{h^{-1}\,{\rm kpc}}
\usepackage{graphicx}
\usepackage{dcolumn}
\usepackage{bm}

\begin{document}

\preprint{PUPT-2145}

\title{Structure Formation With a Long-Range\\ Scalar Dark Matter Interaction}

\author{ Adi Nusser}
\affiliation{Physics Department and the Asher Space Research Institute, Technion, Haifa 32000, Israel}

\author{S. S. Gubser}
\affiliation{Joseph Henry Laboratories, 
Princeton University,
Princeton, NJ 08544, USA}

\author{P. J. E. Peebles}
\affiliation{Joseph Henry Laboratories, 
Princeton University,
Princeton, NJ 08544, USA}

\begin{abstract}
Numerical simulations show that a long-range scalar interaction in a single species of massive dark matter particles causes voids between the concentrations of large galaxies to be more nearly empty, suppresses accretion of intergalactic matter onto galaxies at low redshift, and produces an early generation of dense dark matter halos. These three effects, in moderation, seem to be improvements over the $\Lambda$CDM model predictions for cosmic structure formation. Because the scalar interaction in this model has negligible effect on laboratory physics and the classical cosmological tests, it offers an observationally attractive example of cosmology with complicated physics in the dark sector, notably a large violation of the weak equivalence principle. 
\end{abstract}

\pacs{98.80.-k, 
11.25.-w, 
95.35.+d, 
98.65.Dx 
}
\maketitle

\section{Introduction}
\label{sec:Introduction}

The $\Lambda$CDM model (cold dark matter with a cosmological constant) passes demanding observational tests, but it may be possible to improve the agreement with observations by introducing an additional long-range interaction in the dark sector.  In this paper, we consider a force law between two dark matter particles, each of mass $m$, arising from the potential
 \beq
  V = -{G m^2 \over r} \left( 1 + \beta e^{-r/r_s} \right) \,,
 \label{eq:V}
 \eeq
where $\beta$ is of order unity and positive \footnote{A gauge interaction between identical particles would give negative $\beta$.} and the screening length $r_s$ is on the order of $1\,{\rm Mpc}$ today and constant in co-moving coordinates.  This type of force law, for particles at a separation $r$ much less than the Hubble scale, can arise from ordinary Newtonian gravity plus a massless scalar, where the scalar couples to the dark matter particles and to an additional relativistic particle species whose dynamics generates the screening length $r_s$. The couplings of the scalar to the visible sector must be minimal in order not to conflict with laboratory tests of Newton's law: the proposed modification (\ref{eq:V}) applies only to dark matter particles. 

The idea of a long-range scalar interaction has been under discussion for a long time, in many contexts, as reviewed in \cite{BransSummary:1997,fp:2003}. The idea that the scalar interaction may couple only dark matter particles was introduced in \cite{Damour:1990tw}. The field may relax to near minimum of its potential energy expressed as a sum over particle masses. In the example considered in \cite{Damour:1994zq} this can eliminate the scalar interaction. In \cite{Gradwohl:SDG}  the scalar has a mass term with minimum offset from the minimum of the dark matter particle mass as a function of the scalar field. This produces a scalar interaction of the form (\ref{eq:V}) with $r_s$ a fixed proper length. The first analyses of the effect on dark mater halo formation and interactions are in \cite{Gradwohl:SDG,Gradwohl:1992ue,Casas:1991ky}. 
In the model discussed here the scalar field finds its minimum potential energy where the screening particles have close to zero mass, with the result that the cutoff length $r_s$ among the nonrelativistic dark matter species expands with the general expansion of the universe. This approach, which is developed in \cite{fp:2003,gpOne:2004,gpTwo:2004}, offers a variety of scenarios for cosmology: there may be several scalars with various couplings to several species of non-relativistic dark matter.  For the purpose of a numerical exploration of what this approach can offer to physical cosmology, we focus here on a model with one non-relativistic dark matter species and one relativistic screening species. 

With $\beta \sim 1$, $r_s \sim 1\,{\rm Mpc}$ today, and $r_s$ constant not as a physical length but in co-moving coordinates, there is little effect on the classical cosmological tests, provided we assume that initial conditions are adiabatic and that the coupling to the screening particles suppresses evolution of the mean scalar field value.  But our numerical simulations show that the effect on structure formation on the scale of galaxies is pronounced.  The scalar force lowers the density of dark matter in the present-day voids between the concentrations of large galaxies.  It suppresses the rate of accretion of intergalactic debris onto galaxies at low redshift.  And it increases the redshift of assembly of dark matter halos with comoving sizes smaller than $r_s$.

One could reproduce some of these arguably attractive results in a $\Lambda$CDM model with the initial mass distribution contrived to match the present distribution of mass in our scalar interaction model.  The match could be only approximate, of course, because the scalar force not only speeds up formation of galaxy-size halos but also also affects their interaction.  And, in standard $\Lambda$CDM, 
it seems difficult to generate early haloes without also having substantial late time accretion. 

In \S\ref{sec:Dynamics} we review the dark matter dynamics in the scalar field model under consideration and the motivation for the model from string theory.  In \S\ref{sec:Structure} we present the results of numerical simulations of structure formation.  Our simulations have modest spatial resolution and numbers of particles, compared to what can be done, and should therefore be regarded as only a first exploration of what the model has to offer.  Nevertheless, as discussed in section~\S\ref{subsec:Observations}, we can conclude that the effects we do establish are observationally attractive.  In \S\ref{sec:Halos}, we comment on issues related to massive halos which we cannot analyze using our simulations, but which we hope will be explored in future work.

\section{Dynamics}
\label{sec:Dynamics}

The form (\ref{eq:V}) is identical \footnote{In the fifth force literature, the strength and range parameters are usually denoted $\alpha$ and $\lambda$.} to the standard parametrization of hypothetical fifth force corrections to Newton's law of gravity in the visible sector (see for example the review article \cite{Adelberger:2003}).  The interesting regime for the strength parameter $\beta$ is also the same as in discussions of a fifth force: $\beta \sim O(1)$.  But the interesting scale for the screening length in our case is $r_s \sim 1\,{\rm Mpc}$, vastly larger than the regime $r_s \sim 100\,{\rm \mu m}$ of interest in modern fifth force experiments.  And, crucially, we assume that the scalar force acts only in the dark sector: visible matter interacts with the dark matter only through ordinary gravity.

It is essential that we assume that the self-interaction potential $V(\phi)$ for the scalar field can be neglected: a significant non-zero $V'(0)$ would drive $\phi$ away from $0$ at late times, and non-zero $V''(0)$ (with $V'(0)=0$) amounts to a mass for $\phi$ which dominates over the screening effect of the relativistic particles at late times.  Non-zero higher derivatives of $V$ at $\phi=0$ might endanger our story in subtler ways which have not been fully probed. These requirements on $V(\phi)$ are in conflict with standard notions of field-theoretic naturalness.  We nevertheless claim some motivation for our model from string theory and supersymmetry, as we briefly review in \S\ref{subsec:StringTheory}.

Our further discussion in sections~\ref{subsec:Screening} and~\ref{subsec:StringTheory} of the physics behind the force law (\ref{eq:V}) includes some recapitulation of earlier discussions \cite{fp:2003,gpOne:2004,gpTwo:2004}, presented here in the interest of a self-contained presentation.

\subsection{The screening mechanism}
\label{subsec:Screening}

Let one dark matter species have mass $m-y\phi$, while a second species has mass $y_s\phi$.  The couplings $y$ and $y_s$ are positive dimensionless constants.  Assume that the particles of mass $y_s\phi$ have a much larger number density, $\bar n \ll \bar n_s$, so that as the field moves to minimize the energy in particle masses it is pulled to $0<\phi\ll m$.  That makes the screening particles relativistic and gives the dark matter particles masses that are close to $m$. The particle dynamics, independent of spin or statistics, can for present purposes be modeled as gasses of classical point-like particles with the action
\beq
S = \int d^4x \, {1 \over 2} (\partial\phi )^2 - 
 \sum_\alpha\int_{\gamma_\alpha} ds \, m_\alpha(\phi) \,,
 \label{eq:action}
\eeq
where the sum is over all the particles, and $m_\alpha(\phi) = m-y\phi$ or $y_s \phi$ depending on the species of particle $\alpha$.

During structure formation the scalar field dynamics can be treated in a quasi-stationary approximation:
\beq
\nabla ^2\phi = \phi /r_s^2 - yn({\bf r},t)\,, \label{eq:fieldeq}
\eeq
where $\nabla^2$ is the spatial laplacian and
\beq
r_s = \sqrt{\epsilon_s/y_s^2\bar n_s}\,.
\label{eq:rs}
\eeq
The last term in (\ref{eq:fieldeq}) accounts for the non-relativistic particles in a hydrodynamic approximation.  The previous term, $\phi/r_s^2$, follows by noting \cite{fp:2003} that the source term for $\phi$ for a particle with speed $v$ includes a factor $ds/dt = \sqrt{1-v^2}$, and that for quasi-static configurations of $\phi$ the screening particle energy, $\epsilon_s = m_s/\sqrt{1-v^2} = y_s \phi/\sqrt{1-v^2}$, is nearly independent of position.  Elimination of $\sqrt{1-v^2}$ in favor of $\epsilon_s$ results in the screening length (\ref{eq:rs}).  The energy $\epsilon_s$ does change with time, scaling with the expansion of the universe as $\epsilon_s\propto a(t)^{-1}$.  The screening length thus scales as $r_s\propto a(t)$, that is, the comoving length is constant.

The scalar field produced by a single dark matter particle at distance $r\ll r_s$ is $\phi = y/4\pi r$. The force this field exerts on another dark matter particle is the negative of the gradient of the mass $m-y\phi$, that is, ${\bf F} = y\nabla\phi$, so we see that the particles are attracted with force $F=y^2/4\pi r^2$ at $r\ll r_s$. This means the ratio of the scalar and gravitational forces of attraction of two dark matter particles is
\beq
\beta = {y^2\over 4\pi Gm^2}. \label{eq:beta}
\eeq
The relations (\ref{eq:rs}) and (\ref{eq:beta}) summarize how the parameters of the potential (\ref{eq:V}) emerge from the dynamics we have added to the dark sector.

The effect of the scalar interaction on the evolution of the  mass density contrast $\delta=\delta\rho/\rho$ in linear perturbation theory is simply expressed in terms of the Fourier amplitudes $\delta _{\bf k}(t)$. In the approximation that all the mass is in the dark matter, the evolution equation is
\beq
\ddot\delta _{\bf k} + 2{\dot a\over a}\dot\delta_{\bf k}=4\pi G\bar\rho
\left[1+\frac{\beta}{1+(k r_{s})^{-2}}\right]\delta _{\bf k},
\label{eq:lin}
\eeq
where $k$ and $r_s$ are constant in comoving coordinates. In the Einstein-de Sitter model, modes with  wavenumber $k$ grow as
\beq
\delta _{\bf k}\propto t^p,\qquad p=\frac{1}{6}{\left[{25+\frac{24\beta}{1+(k r_{s})^{-2}}}\right]^{1/2}-\frac{1}{6}},
\label{eq:p}
\eeq
to be compared to the usual power law,  $p=2/3$, at $kr_s\ll 1$. The scalar interaction thus causes earlier development of small-scale structure. We comment on the possible implications in \S\ref{sec:Halos}.

The value of the interaction cutoff length $r_s$ is limited by the allowed energy density $\rho _s=\epsilon_s\bar n_s$ in the relativistic screening matter. The ratio to the mean mass density is
\beq
{\rho _s\over\rho} = {3\over 2}\beta\Omega (Hr_s)^2S^2,
\label{eq:rhos}
\eeq
where the ratio of mean number densities of screening and dark matter particles is
\beq
S ={y_s\bar n_s\over y\bar n}. \label{eq:Rs}
\eeq
To avoid violating the standard model for the origin of light elements at high redshift we must choose parameters so the mass density in screening matter is less than that of about two low mass neutrino families \cite{Cyburt:2004yc}, or $\rho_s\la 0.5 aT^4$, where the present background radiation temperature is $T=2.725\,{\rm K}$. For the Hubble and matter density parameters
$ H=70\,{\rm km\,s^{-1}\,Mpc^{-1}}$
 and $\Omega=0.3$,
this condition in eq.~(\ref{eq:rhos}) says
\beq
\beta S^2r_s^2\la 3000,
\eeq
where $r_s$ is measured in megaparsecs.

We can write the present screening particle energy as
\beq
{\epsilon_s \over M_{\rm Pl}}
 ={3 \over 2} y_s\beta^{1/2}S\Omega (Hr_s)^2
\simeq 5 \times 10^{-8}y_s\beta^{1/2} S r_s^2\,,
\label{eq:epsilon_value}
\eeq
where $M_{\rm Pl} = 1/\sqrt{8\pi G} \approx 2.4 \times 10^{18}\,{\rm GeV}$ is the reduced Planck mass.

The growing concentrations of dark matter pull $\phi$ away from zero. Where $y_s\phi$ exceeds $\epsilon _s$ the screening particles are excluded --- their mass would exceed their energy --- and the scalar force is unscreened. The condition for this to happen is most simply stated in the limit where the characteristic width of the concentration is greater  than $r_s$, so $\nabla ^2\phi$ is small  compared to $\phi /r_s^2$. Then eq.~(\ref{eq:fieldeq}) may be rewritten as 
\beq
1 +\delta = S\langle\sqrt{1-v^2}\rangle < S \,,
\eeq
where the matter density contrast is $\delta=n/\bar n-1$. When the length scale of the dark matter concentration is smaller than $r_s$ the screening matter is excluded from regions with a larger value of $\delta$. Our numerical simulations ignore this exclusion of screening particles, so we are underestimating the the scalar attraction in regions where the dark matter mass density is large.  Since the screening lengths we shall consider are greater than the sizes of the regions where the dark matter density is large enough to exclude the screening particles our simulations can ignore this somewhat complicated situation and focus on a spatially uniform $r_s$ which is constant over time in comoving coordinates.

\subsection{String theory motivations}
\label{subsec:StringTheory}

In the string theory literature, there is an extended development, starting with \cite{Kripfganz:1988,Brandenberger:1989} and continuing with substantial works on ``brane-gas cosmology'' \footnote{Because our work overlaps only slightly with the existing work on brane-gas cosmology to date, we feel justified in citing two recent papers \cite{Easther:2003,Watson:2003} from this literature whose list of references can serve as a guide to further exploration.}, of possible cosmological implications of strings or branes wrapping cycles in extra dimensions whose sizes and shapes are described by massless scalars in four dimensions.  For the most part, this literature is concerned with effects that are uniform over the three spatial dimensions that we observe.  For example, finite uniform number densities of strings with momentum or winding number around a circular fifth dimension tend to stabilize the size of the circle \cite{Brandenberger:1989}, because the masses of the winding and momentum strings are convex functions of the canonically normalized scalar controlling the size of the circle.

The global picture emerging from this literature is the following.  At tree level, most known compactifications of string theory have a number of moduli (massless scalars) and a number of stable, heavy, point-like objects whose ten-dimensional description is in terms of wrapped or stretched branes (or, sometimes, Kaluza-Klein excitations).  The masses of these objects are functions of the scalars, and these functions typically vanish each at a different point in moduli space.  Sometimes (as for strings wrapping a circle) the vanishing point is at an infinite distance in moduli space.  Tree-level exchanges of the massless scalars mediate long-range forces in four dimensions that are comparable to the force of gravity: in our terminology, $\beta \sim O(1)$.  The reason is that these scalars are themselves gravitons (or perhaps superpartners of gravitons), but with polarization tensors oriented in the extra dimensions.  There are also gauge interactions between the wrapped branes whose strength is roughly gravitational and whose ten-dimensional origins include gravitons whose polarization tensor has one index in the compact directions and one in the four that we observe.  But gauge interactions will be of little interest to us because their effect on structure formation is less pronounced than scalar interactions of comparable strength \footnote{Gauge interactions cause like charges to repel, whereas scalar interactions cause them to attract.  So, in the linear regime, an isocurvature mode describing separation of positive and negative gauge charges grows less quickly than the adiabatic mode.  This is in contrast to the growth (\ref{eq:p}) caused by scalar interactions, which is faster than in standard $\Lambda$CDM.  Likewise, in the non-linear regime, the main dynamical effect of a gauge interaction would be to maintain approximate local charge neutrality, whereas a scalar force actually adds to the gravitational tendency to increase the density of overdense regions.}.

This picture clearly presents a rich and fascinating field of possibilities for dark sector physics, particularly when we contemplate string compactifications with several dozen complex scalar moduli and dozens if not hundreds of stable brane configurations.  But there are some potential problems too:
 \begin{enumerate}
  \item After supersymmetry is broken, a potential is typically generated for all the moduli on at least the ${\rm meV}$ scale.  
  \item Even before supersymmetry breaking, there typically are couplings of the scalars to the visible sector which would violate fifth-force constraints---unless said scalars have a mass greater than roughly an ${\rm meV}$.
  \item If we want the dark matter to be wrapped branes, some mechanism must be specified to generate the right number density of them at early times.
 \end{enumerate}
To implement a screening mechanism in a cosmologically interesting range of parameters (in particular, $\beta \sim 1$ and $r_s \sim 1\,{\rm Mpc}$), the typical energy $\epsilon_s$ (\ref{eq:epsilon_value}) of a screening particle should be on the order $10^{-7} y_s M_{\rm Pl}$ today.  Arranging for such a large per-particle energy could be added to point~3 in the above list as a potential problem: ordinary production mechanisms of light particles seem unlikely to lead to large $\epsilon_s$.

In \cite{gpOne:2004,gpTwo:2004}, some suggestions were made about how points~1 and~2 might be addressed.  Before supersymmetry breaking, point~2 is mostly an issue of controlling the Kahler potential $K(\phi^\dagger,\phi)$ for the scalars which remain massless.  It is sufficient for the Kahler potential to have a minimum with respect to these scalars at the point in moduli space where we sit today.  Then scalar couplings to the visible sector will start at dimension six, and estimates \cite{gpOne:2004} show that solar system tests of Newton's law of gravity are not compromised (and table-top laboratory tests would be entirely unaffected).  It seems that we are asking rather a lot when we demand that the Kahler potential should be quadratic in the moduli just when one of the dark matter species becomes massless.  But, as discussed in \cite{gpTwo:2004}, in certain supersymmetric theories with non-abelian gauge fields and adjoint matter, the gauge invariance forbids linear terms in the Kahler potential along the Coulomb branch of the moduli space: precisely what we want.  In such a situation, the massless scalar field $\phi$ (arising from the adjoint matter) would be complex, the screening particles would include the non-abelian gauge bosons, and the whole ensemble could be arranged to be weakly coupled in the infrared by including enough matter fields (for instance, fundamentals and anti-fundamentals).  It seems very plausible that some suitably constructed set of coincident or intersecting branes could realize the physics outlined here, though an explicit example that includes a non-relativistic species with suitable couplings would be desirable.

Clearly, the sticking point is supersymmetry breaking.  Even if one assumes that the fundamental scale of supersymmetry breaking is $\Lambda_{\rm SUSY} = 10\,{\rm TeV}$ and that its effects are felt in the dark sector only through gravitational mediation, it is still difficult in conventionally understood scenarios to push scalar masses significantly below the gravitino mass, which is roughly
 \beq
  m_{3/2} \sim {\Lambda_{\rm SUSY}^2 \over M_{\rm Pl}} \sim 0.1\,{\rm eV}
   \,.
  \label{eq:GravitinoMass}
 \eeq
Without supersymmetry, the expectation from naturalness is that scalar masses would be much bigger.

The estimate (\ref{eq:GravitinoMass}) is the cleanest naturalness argument against our proposal.  There are, however, some reasons to believe it can be circumvented.  In the duality between strings in anti-de Sitter space (AdS) and a conformal field theory (CFT) on its boundary \cite{Maldacena:1997,gkp:1998,Witten:1998}, the existence of Lorentz-invariant operators with dimension $\Delta \sim O(1)$ in the CFT guarantees that there is a scalar field in AdS with mass 
 \beq
  m = \sqrt{\Delta(\Delta-4)}/L \,,
  \label{eq:adsM}
 \eeq
where $L$ is the radius of curvature of AdS---the analog of the Hubble scale.  Supersymmetry is not invoked in any way in this argument, though the conclusion would come as a surprise to proponents of naturalness when $L$ is much bigger than the string scale.  It is conjectured \cite{VafaConjecture} that similar arguments play out in the setting of an expanding universe entering a de Sitter phase, with the conclusion that there should be scalars with Hubble scale masses.  The constraints on possible interactions of such scalars with dark matter are not well understood by the present authors, but they are presumed to be less severe than for Goldstone bosons.

Also, there have been suggestions that scalar masses may be made extremely small if strong coupling renders mass terms irrelevant \cite{ThomasConjecture}, though there is again some question of what interactions are possible for such scalars.

The need for highly energetic screening particles (point 3 above) may prove to be the thorniest issue.  Large $\epsilon_s$ might be arranged if the screening particles are the decay product of some very heavy and moderately long-lived particle---say a Planck mass particle that decays at $z=10^7/y_s$.  It should also be recalled from \cite{gpOne:2004,gpTwo:2004} that one may assume more moderate values of $\epsilon_s$ and the mass $m$ of the non-relativistic particles---bringing the problem of abundances perhaps more within reach of standard ideas like thermal production and freeze-out---if one lowers $y_s$ and $y$ proportionally, so as to preserve $\beta$ and $r_s$.

The simple model (\ref{eq:action}) with one non-relativistic species and one relativistic one might arise straightforwardly as one characteristic behavior of many different underlying string constructions: the relativistic species corresponds to the wrapped branes that happen to be most numerous in the initial conditions; the non-relativistic species corresponds to the next most numerous type of wrapped branes; and other massive species comprise negligible number and energy densities.

In summary, if certain favorable features of classical string theory or unbroken supersymmetry are preserved at low energies in the dark sector, then the scalar interactions we study may arise from a broad variety of string compactification scenarios.  Thus, it is just possible that some simple ideas from string theory are closer to astrophysical observation than has previously been appreciated.

\section{Numerical Simulations of Structure Formation}
\label{sec:Structure}

We turn now to exploring the consequences of the scalar interaction 
on the formation of structure in the expanding universe. This section deals with numerical simulations of structure formation on scales $\sim 100$~kpc to $\sim 5$~Mpc. We comment on the demanding issue of structure formation on smaller scales in \S\ref{sec:Halos}.

\subsection{Numerical Methods}

The treatment of the particle motions given their peculiar  accelerations is standard. To simplify notation in the computation of accelerations we write the scalar field as
\beq
\Phi = -4\pi G m\phi /y.
\eeq
Then the force on a dark matter particle is 
\beq
{\bf F} = - m\nabla\Phi _g + y\nabla\phi = -m\nabla (\Phi _g +\beta\Phi ),
\eeq
where $\Phi _g$ is the Newtonian gravitational potential and $\beta$ is the ratio of the scalar to gravitational forces on scales small compared to $r_s$ (eq.~[\ref{eq:beta}]). In this notation the Fourier transforms of the gravitational and scalar field equations are 
\beq
k^2\Phi _g({\bf k}) = (k^2+r_s^{-2})\Phi ({\bf k})=-4\pi G\bar\rho\delta ({\bf k}),
\label{eq:fourier}
\eeq
where $\bar \rho= m\bar n$ is the dark mass density, $\delta=\rho/{\bar \rho}-1$ is the local density contrast.
We have neglected the mass in baryons and   assumed that the spatial average of $\nabla^{2}\phi$ is zero. 
An FFT particle accelerator is easily adjusted from the standard computation of the gravitational potential $\Phi _g$ to compute the scalar field $\Phi$. 

To get some indication of the baryon distribution we follow the motion of  trace particles under the gravitational acceleration ${\bf g} = -\nabla\Phi _g$ alone. These particles have the same number and initial
conditions as the dark matter.
 We call these particles ``baryons'' though their motions do not take account of  hydrodynamical forces. 

The particle motions  were simulated with
a particle-mesh (PM) FFT N-body code kindly provided by E. Bertschinger \cite{bertgelb}. All simulations use $128^3$ particles in a cubic box
of $128^{3}$ grid points for FFT computations. The initial conditions are represented by displacements from a cubic lattice to represent Gaussian density fluctuations with the $\Lambda$CDM power spectrum normalized so that the 
linear value of the rms density fluctuations in spheres of radius $8\hmpc$
is $0.9$ at the present time.  (The Hubble parameter is $H = 100h\,{\rm km\,s^{-1}\,Mpc^{-1}}$.)  All runs started at redshift $z=100$.
The N-body code is run in practice
with $\Omega = 0.3$ and $\Lambda=0$, but the output times are scaled to $\Omega=0.3$ and $\Lambda=0.7$ using the recipe in \cite{nussercolberg}.  

\subsection{Simulations}

We present results from simulations with two values of the box widths, $L=10$
and $50\hmpc$. For a given set of the initial conditions,
simulations were run for several values of  $\beta $ and $r_{s}$.
For the larger box size $L=50\hmpc$ we obtained six simulations 
with different initial phases for  each pair  of $\beta$ and $r_{s}$.
For $L=10\hmpc$ we ran two simulations with different initial phases.

The main theme in the following discussion is the effect of the scalar interaction in suppressing the amount of dark matter debris between and around the massive dark matter halos. This is illustrated first in two maps of particle distributions and then in some statistics.  The observational appeal is discussed in \S\ref{subsec:Observations}.

\begin{figure}
\hspace{-0.45cm}
\includegraphics[width=3.5in]{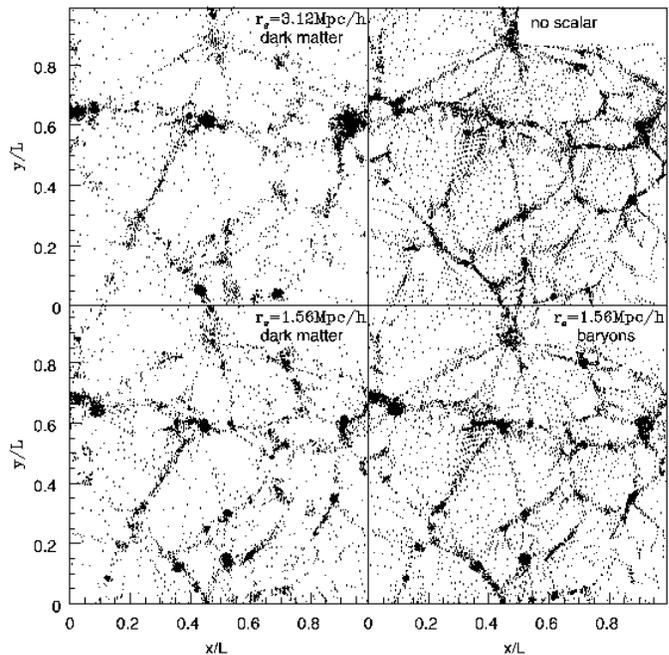}
\vspace{-0.1cm}
\caption{Present particle distributions in a slice $0.5\hmpc$ deep through a simulation box of width $50\hmpc$. The initial conditions are the same in each simulation. The panel at the top right shows the distribution in the absence of the scalar force. In the other simulations $\beta =1$. The baryons in the lower right panel respond to the mass distribution in the lower left panel. 
\label{fig:xy}}
\end{figure}

Figure~\ref{fig:xy} shows the effect of a scalar force of attraction
 with $\beta =1$ for $r_s=1.56\hmpc$ and $3.12\hmpc$ in simulations with 
 box size $L=50\hmpc$. The control case with no scalar force is shown in the slice in the upper right, and the dark matter distributions for the same slice with the same initial conditions and the two values of $r_s$ are shown in the left panels. One sees that the scalar interaction produces more prominent massive halos and leaves less dark matter between the halos and filaments. Since the particles representing baryons respond only to the gravitational force, it is not surprising that the baryon distribution in the lower right panel is less tightly clustered than the mass  distribution in the lower left that produced it and more strongly clustered than the mass distribution in the control plot directly above it. 

\begin{figure}
\hspace{-.3cm}
\includegraphics*[width=3.5in]{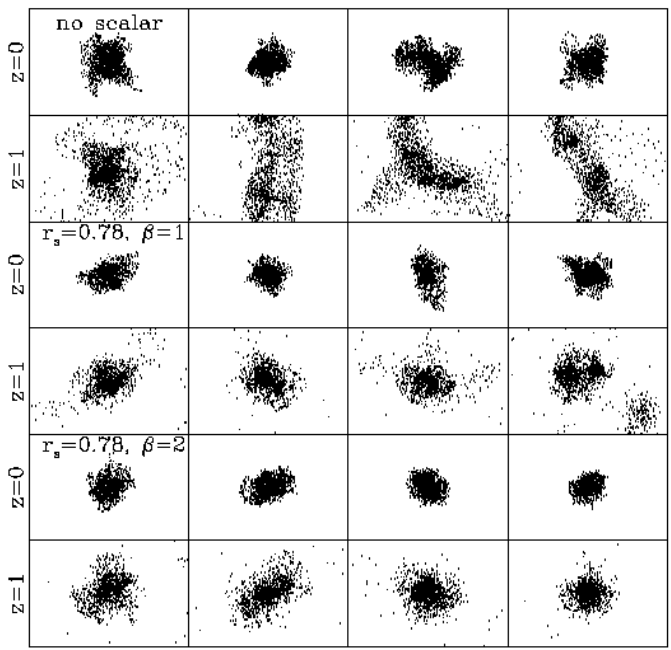}
\vspace{-0.1cm}
\caption{Illustrations of the suppression of accretion at low  redshift. 
 Each panel at $z=0$ shows the projected positions  in a halo selected by the 
friends-of-friends linking algorithm with a linking length of $0.2$ times
the mean particle separation. The panels labeled $z=1$ show the positions at that epoch of the particles selected in the linking list at $z=0$. The comoving width of each panel is $3\hmpc$ and its height is $2\hmpc$. 
\label{fig:merger}}
\end{figure}

Another aspect of the suppression of debris between massive halos is illustrated in figure~\ref{fig:merger}, which shows examples of the late time evolution of halos with and without the scalar interaction. For the purpose of our analysis, a ``massive halo'' is identified 
by the standard friends-of-friends (FOF) algorithm.  In this algorithm, a particle is in a group if it is within a prescribed linking length of some other particle in the group.  The simulation box width for the identification of these halos is $L=50\hmpc$, the mean interparticle separation is $w=L/128$, and the linking length in the FOF algorithm is $0.2w=80\hkpc$.
 The comoving width of each panel is $3\hmpc$ and its height is 
 $2\hmpc$. The particle mass is $10^{9.7}h^{-1}M_\odot$, and  the halo masses in figure~\ref{fig:merger} are in the range $10^{11.7}$ to~$10^{12.7}h^{-1}M_\odot$. The panels labeled $z=0$ show the present positions of the particles identified by the linking list, and the panels labeled $z=1$ show the positions of the same sets of particles at the earlier epoch. The examples without the scalar interaction show the prominent low redshift accretion predicted by the $\Lambda$CDM cosmology.  Turning on scalar interactions leads to less accretion at low redshift, not because the scalar forces discourage accretion (the opposite is true) but because there is less debris to accrete.  Loosely speaking, accretion finishes earlier when scalar interactions are turned on.

\begin{figure}
\hspace{-0.45cm}
\includegraphics*[width=3.5in]{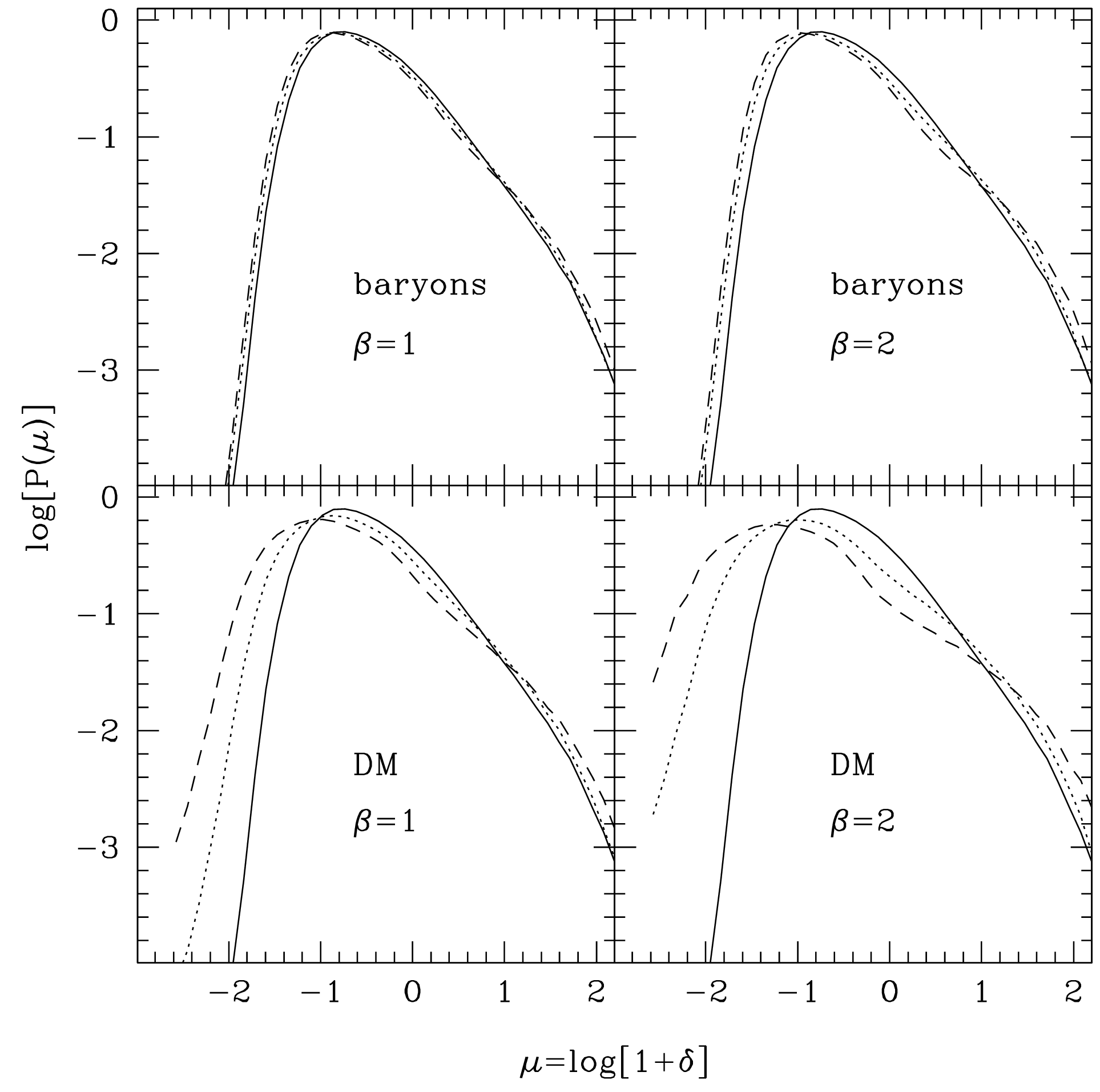}
\vspace{-0.1cm}
\caption{The distributions of the density contrasts in dark matter and baryons smoothed with a top-hat spherical window of radius $1.5\hmpc$ at the present epoch. The standard model is the solid curve, the dotted curve shows the effect of the scalar force with $r_s=0.78\hmpc$, and the dashed curve shows $r_s=1.56\hmpc$. The simulation box width is $50\hmpc$. 
\label{fig:pdf150}}
\end{figure}

\begin{figure}
\hspace{-0.45cm}
\includegraphics*[width=3.5in]{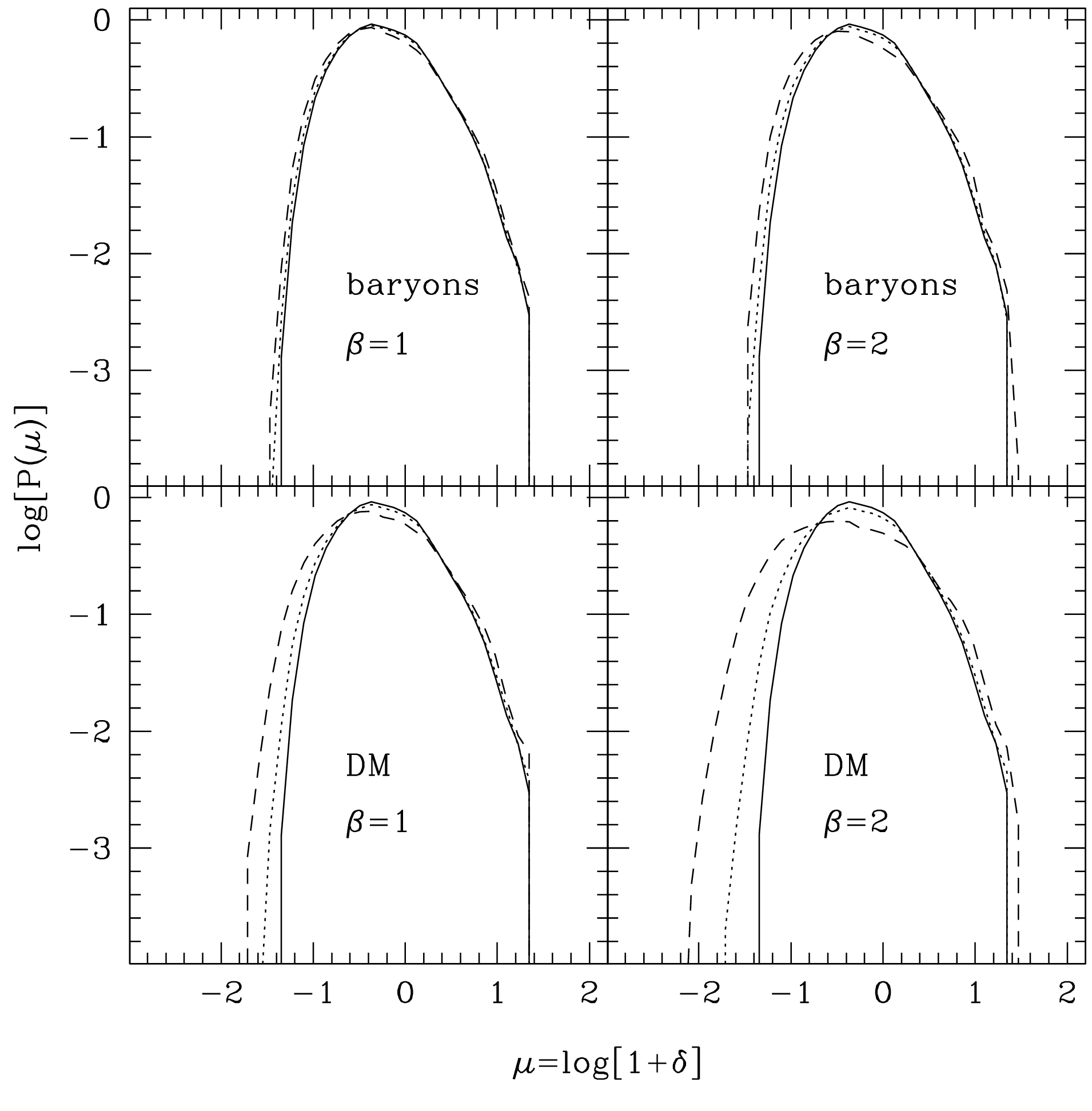}
\vspace{-0.1cm}
\caption{The same as figure~\ref{fig:pdf150} for window radius $5\hmpc$.
\label{fig:pdf500}}
\end{figure}

The behavior illustrated in figures~\ref{fig:xy} and~\ref{fig:merger} is quantified by the probability distribution function in the density contrast found within a randomly placed sphere, which we estimate from  
the density obtained by convolving the  spatial particle distribution  through a spherical window. The sphere radii are $r=1.5\hmpc$ and $r=5\hmpc$ in the distributions in figures~\ref{fig:pdf150} and~\ref{fig:pdf500}. These sample length scales are intermediate between the nominal halo sizes of $L_\ast$ galaxies, $r\sim 300\hkpc$, and the typical sizes of voids, $r\sim 15\hmpc$. The mass distributions for the baryons, shown in the top panels, are not greatly affected by the scalar force, as one sees in figure~\ref{fig:xy}. At both sphere radii the scalar force increases the probability of finding a sphere with large density contrast, but the much larger effect is the increased probability of finding a nearly empty sphere that contains density less than 3\%\ of the mean. 

\begin{figure}
\hspace{-0.3cm}
\includegraphics*[width=3.5in]{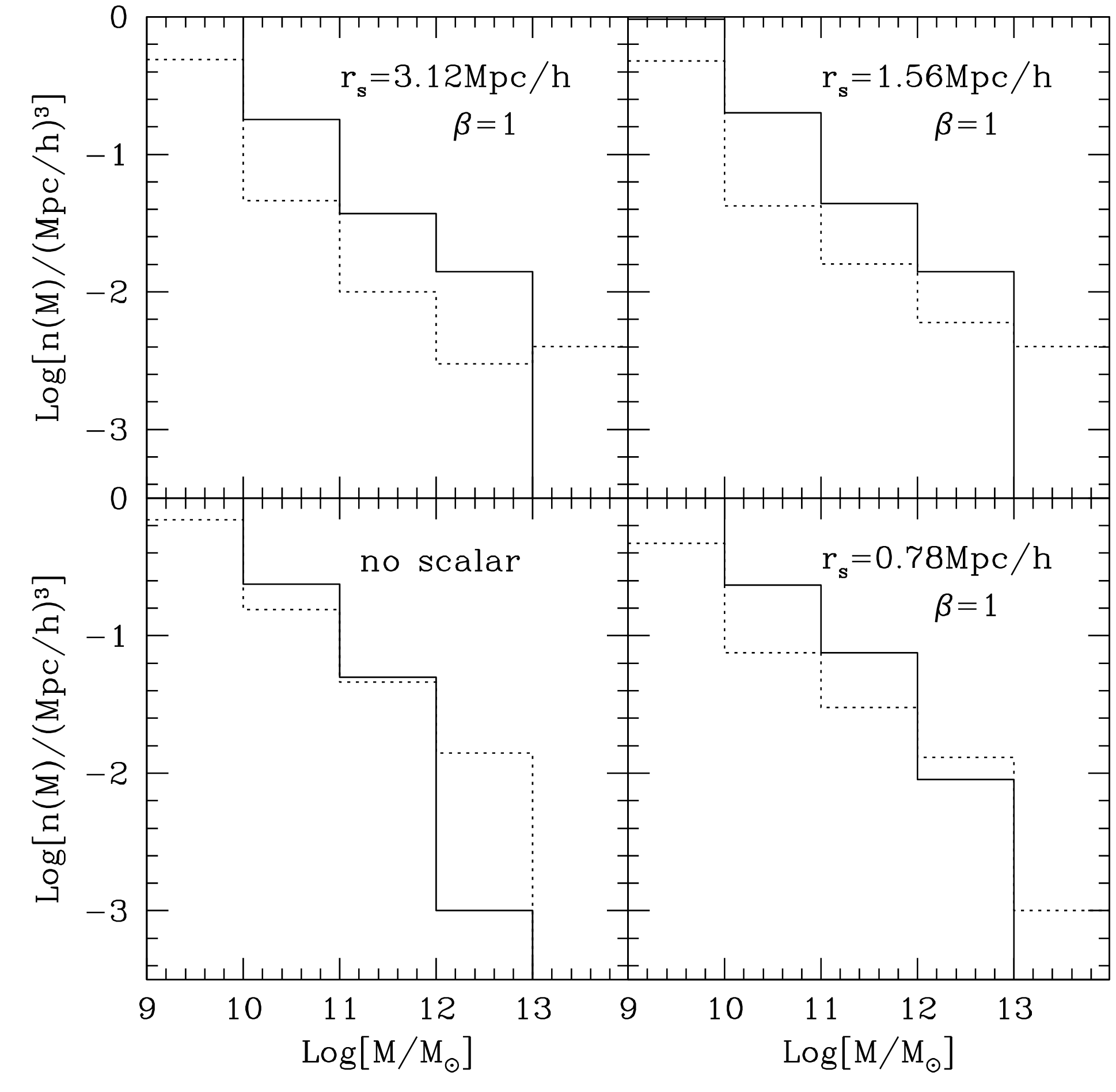}
\vspace{-0.1cm}
\caption{The halo mass function computed in a simulation box with width $10\hmpc$. The dotted line is the present function, the solid  line the mass function at redshift $z=1$.
\label{fig:nm}}
\end{figure}

For a measure of structure formation on smaller scales we use a halo mass function $n(M)$ defined such that 
$n(M)\rm d \log_{10} M $ is the mean number density of halos in the 
relevant mass range. 
Figure~\ref{fig:nm} shows $n(M) $ for halos identified in the $L=10\hmpc$ simulations using the FOF algorithm.  
The linking parameter is $15\hkpc$ and the particle mass is $10^{7.6}h^{-1}M_\odot$. All halos with more than 20 particles 
are included in the calculation of $n(M)$.
Figure~\ref{fig:nm}  shows histograms of the mass function in the simulations with $\beta=1$.
The dotted and solid lines correspond to $n(M)$ at $z=0$ and $z=1$, respectively.
The abundance of halos in the mass range $10^{11.5}-10^{12.5}M_\odot$ 
is consistent with the observed abundance of galaxies thought to  reside 
in halos in this mass range \cite{adelberger2004}. 
The halo mass function can be used to quantify the recent history of merging in halos with $M\sim 10^{12}M_{\odot}$: the scalar interaction shifts merging activities to higher masses relative to the standard dynamics without scalar interaction. 
  This is encouraging as it mitigates some of the problems associated
 with the recent intense merging predicted by standard $\Lambda$CDM  halos of mass
 $\sim 10^{12}M_{\odot}$ (see \S\ref{subsec:Observations}).

The simulations with $L=10\hmpc$ are too small to assess the 
abundance of halos in the mass range corresponding to rich galaxy clusters. 
Therefore we have also computed  $n(M)$ from  the  simulations  with 
$L=50\hmpc$.  
All of these larger simulations yield similar 
prediction for the abundance of cluster  halos. 
This is expected since even though  $r_{s}$ is of the order  
of the virial radii of rich clusters, most of the matter in these objects 
collapsed only recently from much larger comoving distances and therefore its dynamics 
has not been affected by the scalar interaction.
 
\begin{figure}
\hspace{-0.45cm}
\includegraphics*[width=3.5in]{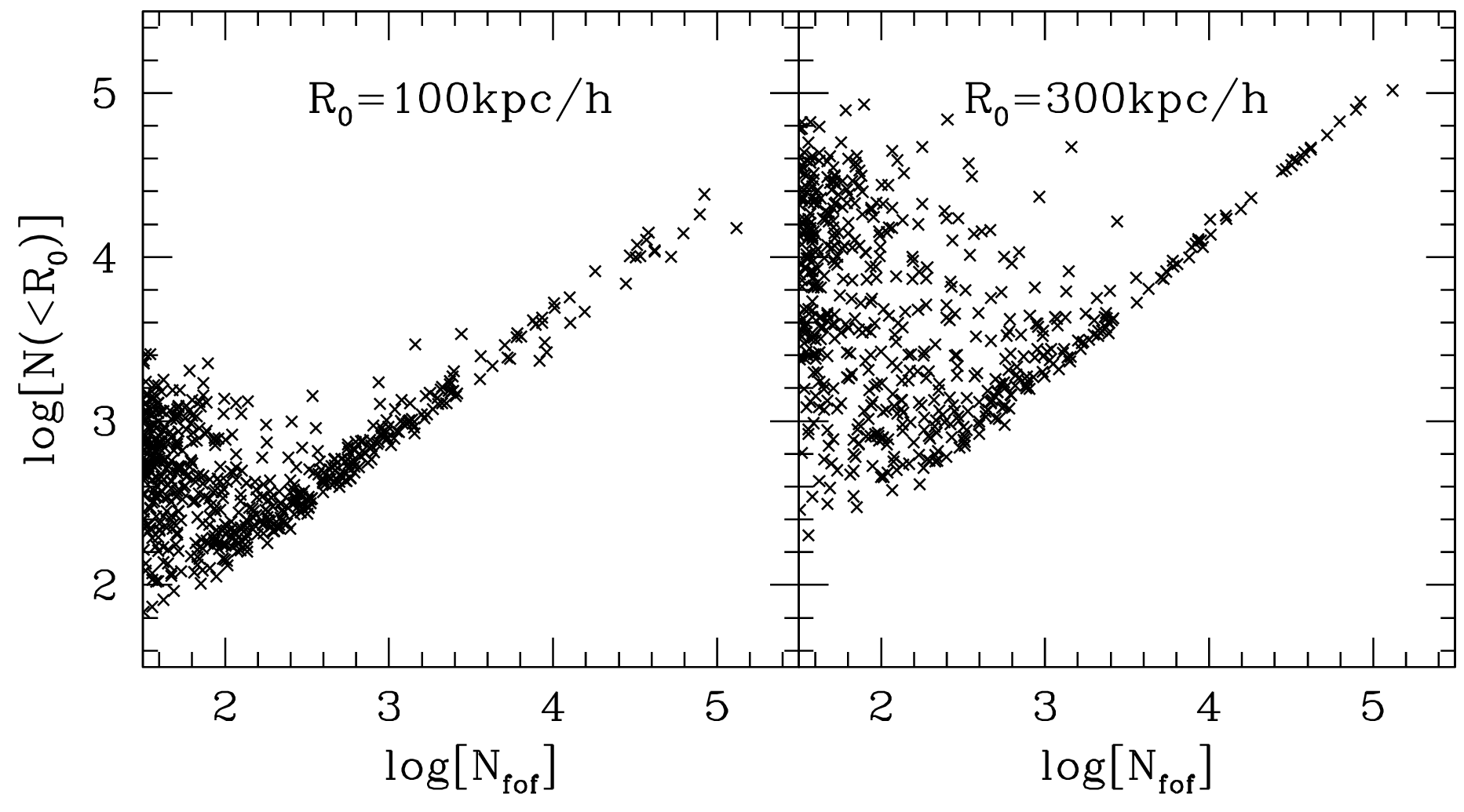}
\vspace{-0.1cm}
\caption{Comparison of the particle number in a halo identified as a linking list with the number of particles within distances 100 and $300\hkpc$ for the simulation with no scalar interaction.
\label{fig:nfof}}
\end{figure}

In light of the limited resolution of our simulations, it is prudent to consider an alternative measure of the halos identified by our FOF algorithm.  For the no scalar simulation, figure~\ref{fig:nfof} shows the numbers of particles $N({<}R_0)$ within radii $R_0 = 100\hkpc$ and $R_0 = 300\hkpc$ of the center of mass of each FOF halo.  This figure gives evidence that the more massive FOF halos have the mass structures and density contrasts of conventional relaxed halos.  The evidence is in two parts.  First, at $N_{\rm fof}>1000$, we see that $N({<}300\hkpc)/N({<}100\hkpc) \approx 3$, indicating agreement with the commonly discussed form $\rho \sim 1/r^2$.  Second, at $N_{\rm fof}=1000$ (corresponding to $M = 10^{10.6} h^{-1} M_\odot$) and $R_0 = 100\hkpc$, the density contrast is $\bar\delta \sim 100$, which is close to the nominal density contrast at virial equilibrium.  For $N_{\rm fof} < 100$, the significant vertical scatter comes from small FOF halos which are close to big ones.  In sum, the behavior illustrated in figure~\ref{fig:nfof} is an encouraging indicator of the robustness of the notion of halos as identified by the FOF algorithm.

\begin{figure}
\hspace{-0.3cm}
\includegraphics*[width=3.5in]{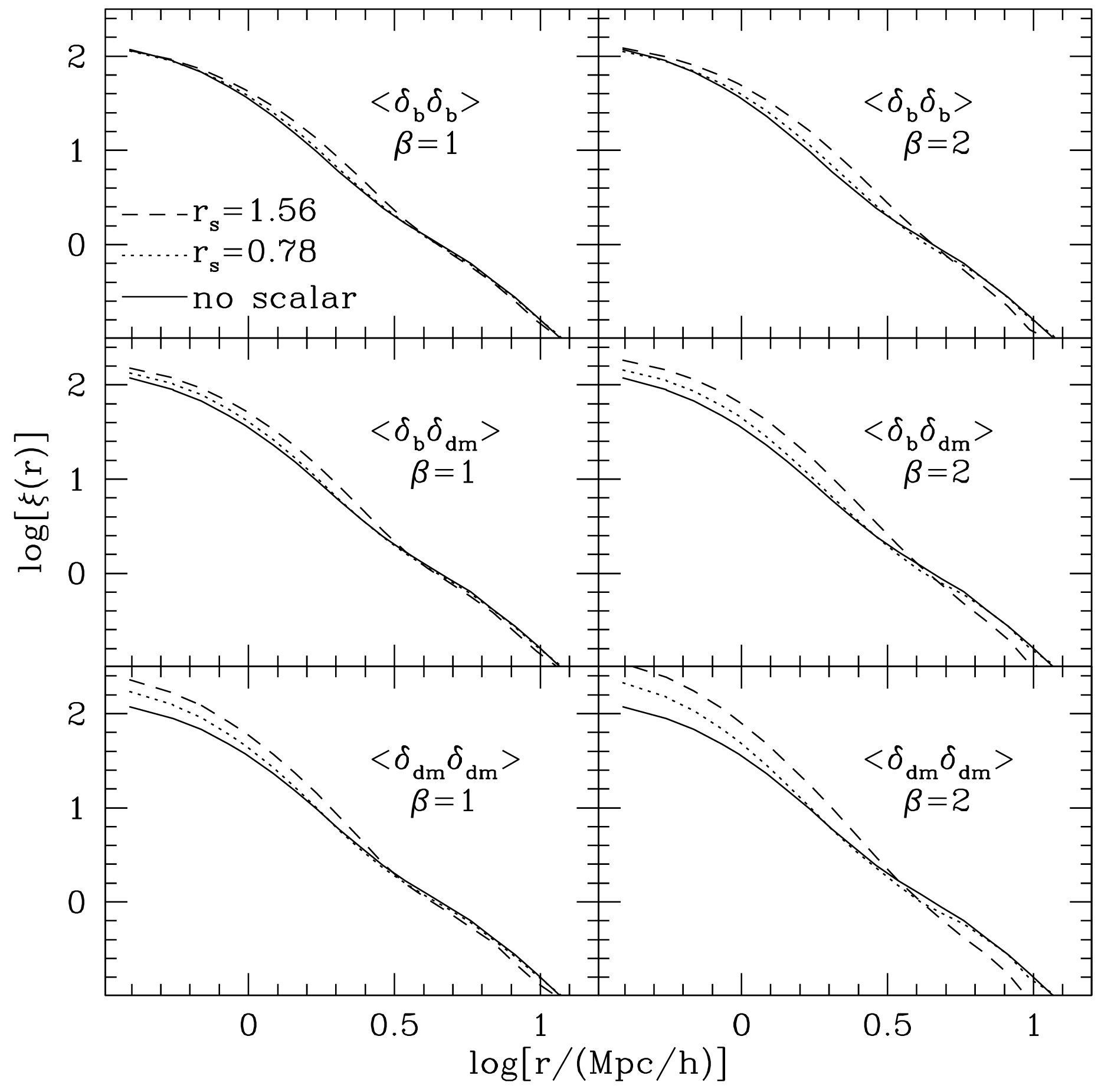}
\vspace{-0.1cm}
\caption{The dark matter and ``baryon'' correlation functions.
\label{fig:corr}}
\end{figure}

Figure \ref{fig:corr} shows the cross- and auto-correlation functions of the present distributions of baryons and dark matter. The bend in the functions at separations $r\la 1$~Mpc is an artifact of the force resolution in the simulation, and should be disregarded, but we nevertheless expect that the effect of the scalar interaction indicated by the differences of the functions at $0.3\la r\la 1$~Mpc is meaningful. The most notable result is the absence of a feature at the length scale $r_s$ of the scalar force, despite the scalar force of attraction at smaller scales.  It is interesting also that at $r\sim 300\hkpc$ this measure of the baryon clustering is not much affected by the scalar interaction. And it is worth noting that the scalar interaction increases the slope of the mass autocorrelation function at 1 to $3\hmpc$. A tantalizing prospect is that this might reduce scale-dependent bias.

\subsection{Observational Situation}
\label{subsec:Observations}

At low redshifts, the scalar interaction lowers the density of dark matter in voids and suppresses the rate of accretion of dark matter by galaxies.  In this section we comment on how these two related effects may improve upon $\Lambda$CDM in matching observations.

Voids between the concentrations of large galaxies contain plasma clouds with atomic hydrogen surface densities $\sim 10^{13}\,{\rm cm^{-2}}$ detected as Ly$\alpha$ resonance absorption lines \cite{void_baryons}, but voids contain strikingly few isolated dwarf or irregular galaxies \cite{PeeblesVoid:2001}. That is not what we might have expected from simulations of the dark mass distribution in the $\Lambda$CDM cosmology, as illustrated in figure~1 of \cite{mwVoid:2002}.  The analysis in \cite{mwVoid:2002} indicates that the distributions of distances to the nearest low mass halo from giant and low mass halos are not very different, consistent with the observations. This is difficult to interpret, however, because the distances in the simulation are an order of magnitude larger than what is found for this statistic applied to galaxy catalogs \cite{PeeblesVoid:2001}, and because our visual impression is that the simulation presented in  \cite{mwVoid:2002} shows distinctly more low mass halos between the concentrations of giants than are observed. This could be because the low mass halos in voids contain too little gas or stars to be observable. But that does not agree with the observations of nearby dwarfs at ambient densities close to the cosmic mean. One would wonder why dwarfs in the voids, where the $\Lambda$CDM cosmology predicts densities just an order of magnitude lower than the mean, are not similarly visible. 

Perhaps the voids appear empty because a scalar force has pushed most of the dark matter out of them. Since the voids would have grown out of smaller density minima (as measured in co-moving coordinates), their growth would have been assisted by the scalar force even at the relatively short ranges $r_s$ we have considered, and indeed figure~\ref{fig:pdf500} shows an appreciable effect on the abundance of very low density regions with diameter $10\hmpc$. In this model it would not be surprising to see HI clouds in the voids, as one sees in figure~\ref{fig:xy}. It will require more detailed analyses to check  whether the parameters can be chosen for consistency with, among other things, the constraints on the mass fraction in the  baryons left in the voids and the observations of empty voids considerably larger than 10~Mpc. 

In the $\Lambda$CDM cosmology one expects considerable accretion at low redshift. This is illustrated in figure~2 of \cite{glpwjMerge:2003}.  Figure~\ref{fig:merger} of the present paper shows that the scalar interaction distinctly suppresses accretion of dark matter at low redshift. The observations seem to favor suppression; we  mention three aspects. First, the evidence reviewed in \cite{Gilmore:2002jv}, from the ages, chemical abundances, and spatial distributions of the stars in our Milky Way galaxy, is that this system has not been substantially disturbed by accretion since redshift $z\sim 1$. Second, Blitz {\it et al.} \cite{blitz} show that if the high velocity HI clouds were falling into the  Milky Way from distances $\sim 1$~Mpc it would be a good match to the accretion expected in the standard $\Lambda$CDM cosmology. However, as discussed in \cite{pisano}, one does not observe HI emission around other galaxies---hence a challenge to $\Lambda$CDM if we accept the interpretation of \cite{blitz}. Third, in the judgement of Abadi {\it et al.} \cite{abadi} the effect of accretion at low redshift on the thin discs of other spiral galaxies is ``not grossly inconsistent with current data''  but ``is worryingly difficult to accommodate within this general scenario.'' We conclude that although the standard cosmology is not ruled out by the observations, it does require that the ordinary-looking Milky Way galaxy is quite unusual, and that the recent accretion by other galaxies is well hidden.  The more straightforward interpretation is that the $\Lambda$CDM cosmology might have to be adjusted, perhaps to include a scalar interaction in the dark sector. 

We have not explored whether the suppression of debris around massive halos is accompanied by a reduction in the numbers of dwarf satellites of a giant galaxy, which arguably would be observationally desirable \cite{cusps}. Checking this requires more detailed simulations.

\section{Massive Halos}
\label{sec:Halos}

With one massive dark matter species the scalar interaction must cause primeval mass fluctuations on scales $\la r_s$ to grow into nonlinear massive halos distinctly sooner than in the standard cosmology. Our simulations are not adequate to show whether this is a challenge for the scalar model, or perhaps an advantage; we can only present some qualitative considerations and suggestions for further work. 

The growth of the first generation of halos commences when the mass density in matter becomes dominant, at redshift $\sim z_{\rm eq}=3500$ for $\Omega = 0.3$ and $h = 0.7$.  The mass density contrast subsequently grows as $\delta\sim a(t)$ on scales larger than $r_s$ and, if $\beta = 1$, as $\delta\sim a(t)^{3/2}$ on smaller scales (eq.~(\ref{eq:p})). This means the growth factor to $z=20$ in the standard model, when significant structure formation commences, is accomplished in this scalar model at redshift $z\sim 100$. 

We get some understanding of the nature of this early generation of massive halos by noting that the Fourier amplitudes of the mass distribution at wavelengths larger than $r_s$ are not much affected by the scalar interaction. This is because the short range of the scalar force gathers dark matter from distances $\sim r_s$, evacuating a compensating hole around a growing halo, so that the gravitational attraction of matter at larger distances is not much different from the standard model. The baryons would tend to lag behind the early growth of the dark matter density contrast in these halos, because the baryons are attracted only by gravity. Thus we expect the formation of a class of gravitationally isolated, compact, baryon-poor halos. The mass autocorrelation functions in figure~\ref{fig:corr} have adequate resolution to be sensitive to the mass in the early halos. We expect that the absence of a feature at $r_s$ is a result of the growth of the clustering of matter gathered from larger scales at lower redshifts. 

More detailed simulations are needed to determine the dark matter mass fraction in these early halos and the accreted baryon mass under the assumption that nongravitational forces may be neglected. It will be exceedingly difficult, but important, to estimate the early star formation rate in these halos, in order to decide whether it can be consistent with the reionization history indicated by the Wilkinson Microwave Anisotropy Probe \cite{WMAP}. Also to be considered is whether the early generation of massive halos helps promote the strikingly early development of the black holes that power the SDSS $z\sim 6$ quasars \cite{Fan}.

\section{Concluding Remarks}
\label{sec:Conclude}

The numerical simulations presented here of a simple version of the scalar dark matter interactions considered in \cite{fp:2003,gpOne:2004,gpTwo:2004} open up a window to the non-linear regime.  It is mainly in this regime, so crucial to comparison with observations, that the scalar force is expected to make a contribution to structure formation.

We have shown that a dynamically screened scalar interaction offers clear relief to two quite troublesome aspects of the $\Lambda$CDM cosmology. First, it promotes lower density of dark matter in the voids. This agrees with the observation that void dwarf and irregular galaxies are rare \cite{PeeblesVoid:2001}. Second, it suppresses accretion of intergalactic debris onto galaxies at low redshifts. This agrees with the evidence that present-day galaxies by and large act like isolated island universes (as discussed, for example, in \cite{Gilmore:2002jv,deBlok:2001xz}. And it is a fascinating possibility that the physics that improves these aspects of the otherwise very successful  $\Lambda$CDM cosmology could be a window to string theory.

Issues remain.  If the scalar interaction preserves standard estimates of the mass density as a function of radius in the relaxed parts of a dark matter halo, then the earlier structure formation on scales $\la r_s$ would increase the halo mass function without affecting the issue of dark matter central cusps in galaxies (as discussed in \cite{cusps}).  But does it?  The critical checks of the halo density run and halo mass function require more extensive numerical studies.

\acknowledgments

We have benefitted from discussions with Neal Dalal and David Hogg and from comments from the referee. We thank  Antonaldo Diaferio for 
providing us with his FOF code.  SSG~was supported in part by the Department of Energy under Grant No.\ DE-FG02-91ER40671, and by the Sloan Foundation.  AN thanks the Princeton Institute for Advanced Study for its hospitality and support.

\bibliography{simulations}

\begin{thebibliography}{35}
\expandafter\ifx\csname natexlab\endcsname\relax\def\natexlab#1{#1}\fi
\expandafter\ifx\csname bibnamefont\endcsname\relax
  \def\bibnamefont#1{#1}\fi
\expandafter\ifx\csname bibfnamefont\endcsname\relax
  \def\bibfnamefont#1{#1}\fi
\expandafter\ifx\csname citenamefont\endcsname\relax
  \def\citenamefont#1{#1}\fi
\expandafter\ifx\csname url\endcsname\relax
  \def\url#1{\texttt{#1}}\fi
\expandafter\ifx\csname urlprefix\endcsname\relax\def\urlprefix{URL }\fi
\providecommand{\bibinfo}[2]{#2}
\providecommand{\eprint}[2][]{\url{#2}}

\bibitem[{\citenamefont{Brans}(1996)}]{BransSummary:1997}
\bibinfo{author}{\bibfnamefont{C.~H.} \bibnamefont{Brans}}
  (\bibinfo{year}{1996}), \eprint{gr-qc/9705069}.

\bibitem[{\citenamefont{Farrar and Peebles}(2004)}]{fp:2003}
\bibinfo{author}{\bibfnamefont{G.~R.} \bibnamefont{Farrar}} \bibnamefont{and}
  \bibinfo{author}{\bibfnamefont{P.~J.~E.} \bibnamefont{Peebles}},
  \bibinfo{journal}{Astrophys. J.} \textbf{\bibinfo{volume}{604}},
  \bibinfo{pages}{1} (\bibinfo{year}{2004}), \eprint{astro-ph/0307316}.

\bibitem[{\citenamefont{Damour et~al.}(1990)\citenamefont{Damour, Gibbons, and
  Gundlach}}]{Damour:1990tw}
\bibinfo{author}{\bibfnamefont{T.}~\bibnamefont{Damour}},
  \bibinfo{author}{\bibfnamefont{G.~W.} \bibnamefont{Gibbons}},
  \bibnamefont{and} \bibinfo{author}{\bibfnamefont{C.}~\bibnamefont{Gundlach}},
  \bibinfo{journal}{Phys. Rev. Lett.} \textbf{\bibinfo{volume}{64}},
  \bibinfo{pages}{123} (\bibinfo{year}{1990}).

\bibitem[{\citenamefont{Damour and Polyakov}(1994)}]{Damour:1994zq}
\bibinfo{author}{\bibfnamefont{T.}~\bibnamefont{Damour}} \bibnamefont{and}
  \bibinfo{author}{\bibfnamefont{A.~M.} \bibnamefont{Polyakov}},
  \bibinfo{journal}{Nucl. Phys.} \textbf{\bibinfo{volume}{B423}},
  \bibinfo{pages}{532} (\bibinfo{year}{1994}), \eprint{hep-th/9401069}.

\bibitem[{\citenamefont{Gradwohl and Frieman}(1991)}]{Gradwohl:SDG}
\bibinfo{author}{\bibfnamefont{B.-A.} \bibnamefont{Gradwohl}} \bibnamefont{and}
  \bibinfo{author}{\bibfnamefont{J.~A.} \bibnamefont{Frieman}},
  \bibinfo{journal}{Phys. Rev. Lett.} \textbf{\bibinfo{volume}{67}},
  \bibinfo{pages}{2926} (\bibinfo{year}{1991}).

\bibitem[{\citenamefont{Gradwohl and Frieman}(1992)}]{Gradwohl:1992ue}
\bibinfo{author}{\bibfnamefont{B.-A.} \bibnamefont{Gradwohl}} \bibnamefont{and}
  \bibinfo{author}{\bibfnamefont{J.~A.} \bibnamefont{Frieman}},
  \bibinfo{journal}{Astrophys. J.} \textbf{\bibinfo{volume}{398}},
  \bibinfo{pages}{407} (\bibinfo{year}{1992}).

\bibitem[{\citenamefont{Casas et~al.}(1992)\citenamefont{Casas, Garcia-Bellido,
  and Quiros}}]{Casas:1991ky}
\bibinfo{author}{\bibfnamefont{J.~A.} \bibnamefont{Casas}},
  \bibinfo{author}{\bibfnamefont{J.}~\bibnamefont{Garcia-Bellido}},
  \bibnamefont{and} \bibinfo{author}{\bibfnamefont{M.}~\bibnamefont{Quiros}},
  \bibinfo{journal}{Class. Quant. Grav.} \textbf{\bibinfo{volume}{9}},
  \bibinfo{pages}{1371} (\bibinfo{year}{1992}), \eprint{hep-ph/9204213}.

\bibitem[{\citenamefont{Gubser and Peebles}(2004{\natexlab{a}})}]{gpOne:2004}
\bibinfo{author}{\bibfnamefont{S.~S.} \bibnamefont{Gubser}} \bibnamefont{and}
  \bibinfo{author}{\bibfnamefont{P.~J.~E.} \bibnamefont{Peebles}}
  (\bibinfo{year}{2004}{\natexlab{a}}), \eprint{hep-th/0402225}.

\bibitem[{\citenamefont{Gubser and Peebles}(2004{\natexlab{b}})}]{gpTwo:2004}
\bibinfo{author}{\bibfnamefont{S.~S.} \bibnamefont{Gubser}} \bibnamefont{and}
  \bibinfo{author}{\bibfnamefont{P.~J.~E.} \bibnamefont{Peebles}}
  (\bibinfo{year}{2004}{\natexlab{b}}), \eprint{hep-th/0407097}.

\bibitem[{\citenamefont{Adelberger et~al.}(2003)\citenamefont{Adelberger,
  Heckel, and Nelson}}]{Adelberger:2003}
\bibinfo{author}{\bibfnamefont{E.~G.} \bibnamefont{Adelberger}},
  \bibinfo{author}{\bibfnamefont{B.~R.} \bibnamefont{Heckel}},
  \bibnamefont{and} \bibinfo{author}{\bibfnamefont{A.~E.}
  \bibnamefont{Nelson}}, \bibinfo{journal}{Ann. Rev. Nucl. Part. Sci.}
  \textbf{\bibinfo{volume}{53}}, \bibinfo{pages}{77} (\bibinfo{year}{2003}),
  \eprint{hep-ph/0307284}.

\bibitem[{\citenamefont{Cyburt et~al.}(2004)\citenamefont{Cyburt, Fields,
  Olive, and Skillman}}]{Cyburt:2004yc}
\bibinfo{author}{\bibfnamefont{R.~H.} \bibnamefont{Cyburt}},
  \bibinfo{author}{\bibfnamefont{B.~D.} \bibnamefont{Fields}},
  \bibinfo{author}{\bibfnamefont{K.~A.} \bibnamefont{Olive}}, \bibnamefont{and}
  \bibinfo{author}{\bibfnamefont{E.}~\bibnamefont{Skillman}}
  (\bibinfo{year}{2004}), \eprint{astro-ph/0408033}.

\bibitem[{\citenamefont{Kripfganz and Perlt}(1988)}]{Kripfganz:1988}
\bibinfo{author}{\bibfnamefont{J.}~\bibnamefont{Kripfganz}} \bibnamefont{and}
  \bibinfo{author}{\bibfnamefont{H.}~\bibnamefont{Perlt}},
  \bibinfo{journal}{Class. Quant. Grav.} \textbf{\bibinfo{volume}{5}},
  \bibinfo{pages}{453} (\bibinfo{year}{1988}).

\bibitem[{\citenamefont{Brandenberger and Vafa}(1989)}]{Brandenberger:1989}
\bibinfo{author}{\bibfnamefont{R.~H.} \bibnamefont{Brandenberger}}
  \bibnamefont{and} \bibinfo{author}{\bibfnamefont{C.}~\bibnamefont{Vafa}},
  \bibinfo{journal}{Nucl. Phys.} \textbf{\bibinfo{volume}{B316}},
  \bibinfo{pages}{391} (\bibinfo{year}{1989}).

\bibitem[{\citenamefont{Maldacena}(1998)}]{Maldacena:1997}
\bibinfo{author}{\bibfnamefont{J.~M.} \bibnamefont{Maldacena}},
  \bibinfo{journal}{Adv. Theor. Math. Phys.} \textbf{\bibinfo{volume}{2}},
  \bibinfo{pages}{231} (\bibinfo{year}{1998}), \eprint{hep-th/9711200}.

\bibitem[{\citenamefont{Gubser et~al.}(1998)\citenamefont{Gubser, Klebanov, and
  Polyakov}}]{gkp:1998}
\bibinfo{author}{\bibfnamefont{S.~S.} \bibnamefont{Gubser}},
  \bibinfo{author}{\bibfnamefont{I.~R.} \bibnamefont{Klebanov}},
  \bibnamefont{and} \bibinfo{author}{\bibfnamefont{A.~M.}
  \bibnamefont{Polyakov}}, \bibinfo{journal}{Phys. Lett.}
  \textbf{\bibinfo{volume}{B428}}, \bibinfo{pages}{105} (\bibinfo{year}{1998}),
  \eprint{hep-th/9802109}.

\bibitem[{\citenamefont{Witten}(1998)}]{Witten:1998}
\bibinfo{author}{\bibfnamefont{E.}~\bibnamefont{Witten}},
  \bibinfo{journal}{Adv. Theor. Math. Phys.} \textbf{\bibinfo{volume}{2}},
  \bibinfo{pages}{253} (\bibinfo{year}{1998}), \eprint{hep-th/9802150}.

\bibitem[{Vaf()}]{VafaConjecture}
\bibinfo{note}{S. Gukov, A. Neitzke and C. Vafa, private communication}.

\bibitem[{Tho()}]{ThomasConjecture}
\bibinfo{note}{S. Thomas, private communication}.

\bibitem[{\citenamefont{Bertschinger and Gelb}(1991)}]{bertgelb}
\bibinfo{author}{\bibfnamefont{E.}~\bibnamefont{Bertschinger}}
  \bibnamefont{and} \bibinfo{author}{\bibfnamefont{J.~M.} \bibnamefont{Gelb}},
  \bibinfo{journal}{Comp. Ph.} \textbf{\bibinfo{volume}{5}},
  \bibinfo{pages}{164} (\bibinfo{year}{1991}).

\bibitem[{\citenamefont{Nusser and Colberg}(1998)}]{nussercolberg}
\bibinfo{author}{\bibfnamefont{A.}~\bibnamefont{Nusser}} \bibnamefont{and}
  \bibinfo{author}{\bibfnamefont{J.~M.} \bibnamefont{Colberg}},
  \bibinfo{journal}{MNRAS} \textbf{\bibinfo{volume}{294}}, \bibinfo{pages}{457}
  (\bibinfo{year}{1998}).

\bibitem[{\citenamefont{Adelberger et~al.}(2004)\citenamefont{Adelberger,
  Steidel, Pettini, Shapley, Reddy, and Erb}}]{adelberger2004}
\bibinfo{author}{\bibfnamefont{K.}~\bibnamefont{Adelberger}},
  \bibinfo{author}{\bibfnamefont{C.}~\bibnamefont{Steidel}},
  \bibinfo{author}{\bibfnamefont{M.}~\bibnamefont{Pettini}},
  \bibinfo{author}{\bibfnamefont{A.}~\bibnamefont{Shapley}},
  \bibinfo{author}{\bibfnamefont{N.}~\bibnamefont{Reddy}}, \bibnamefont{and}
  \bibinfo{author}{\bibfnamefont{D.}~\bibnamefont{Erb}} (\bibinfo{year}{2004}),
  \eprint{astro-ph/0410165}.

\bibitem[{\citenamefont{Stocke et~al.}(2004)\citenamefont{Stocke, Shull, and
  Penton}}]{void_baryons}
\bibinfo{author}{\bibfnamefont{J.~T.} \bibnamefont{Stocke}},
  \bibinfo{author}{\bibfnamefont{J.~M.} \bibnamefont{Shull}}, \bibnamefont{and}
  \bibinfo{author}{\bibfnamefont{S.~V.} \bibnamefont{Penton}}
  (\bibinfo{year}{2004}), \eprint{astro-ph/0407352}.

\bibitem[{\citenamefont{{Peebles}}(2001)}]{PeeblesVoid:2001}
\bibinfo{author}{\bibfnamefont{P.~J.~E.} \bibnamefont{{Peebles}}},
  \bibinfo{journal}{\apj} \textbf{\bibinfo{volume}{557}}, \bibinfo{pages}{495}
  (\bibinfo{year}{2001}), \eprint{astro-ph/0101127}.

\bibitem[{\citenamefont{Mathis and White}(2002)}]{mwVoid:2002}
\bibinfo{author}{\bibfnamefont{H.}~\bibnamefont{Mathis}} \bibnamefont{and}
  \bibinfo{author}{\bibfnamefont{S.~D.~M.} \bibnamefont{White}},
  \bibinfo{journal}{Mon. Not. Roy. Astron. Soc.}
  \textbf{\bibinfo{volume}{337}}, \bibinfo{pages}{1193} (\bibinfo{year}{2002}),
  \eprint{astro-ph/0201193}.

\bibitem[{\citenamefont{Gao et~al.}(2004)\citenamefont{Gao, Loeb, Peebles,
  White, and Jenkins}}]{glpwjMerge:2003}
\bibinfo{author}{\bibfnamefont{L.}~\bibnamefont{Gao}},
  \bibinfo{author}{\bibfnamefont{A.}~\bibnamefont{Loeb}},
  \bibinfo{author}{\bibfnamefont{P.~J.~E.} \bibnamefont{Peebles}},
  \bibinfo{author}{\bibfnamefont{S.~D.~M.} \bibnamefont{White}},
  \bibnamefont{and} \bibinfo{author}{\bibfnamefont{A.}~\bibnamefont{Jenkins}},
  \bibinfo{journal}{Astrophys. J.} \textbf{\bibinfo{volume}{614}},
  \bibinfo{pages}{17} (\bibinfo{year}{2004}), \eprint{astro-ph/0312499}.

\bibitem[{\citenamefont{Gilmore et~al.}(2002)\citenamefont{Gilmore, Wyse, and
  Norris}}]{Gilmore:2002jv}
\bibinfo{author}{\bibfnamefont{G.}~\bibnamefont{Gilmore}},
  \bibinfo{author}{\bibfnamefont{R.~F.~G.} \bibnamefont{Wyse}},
  \bibnamefont{and} \bibinfo{author}{\bibfnamefont{J.~E.}
  \bibnamefont{Norris}}, \bibinfo{journal}{Astrophys. J.}
  \textbf{\bibinfo{volume}{574}}, \bibinfo{pages}{L39} (\bibinfo{year}{2002}),
  \eprint{astro-ph/0207106}.

\bibitem[{\citenamefont{Blitz et~al.}(1999)\citenamefont{Blitz, Spergel,
  Teuben, Hartmann, and Burton}}]{blitz}
\bibinfo{author}{\bibfnamefont{L.}~\bibnamefont{Blitz}},
  \bibinfo{author}{\bibfnamefont{D.~N.} \bibnamefont{Spergel}},
  \bibinfo{author}{\bibfnamefont{P.~J.} \bibnamefont{Teuben}},
  \bibinfo{author}{\bibfnamefont{D.}~\bibnamefont{Hartmann}}, \bibnamefont{and}
  \bibinfo{author}{\bibfnamefont{W.~B.} \bibnamefont{Burton}},
  \bibinfo{journal}{Astrophys. J.} \textbf{\bibinfo{volume}{514}},
  \bibinfo{pages}{818} (\bibinfo{year}{1999}), \eprint{astro-ph/9803251}.

\bibitem[{\citenamefont{Pisano et~al.}(2004)\citenamefont{Pisano, Barnes,
  Gibson, Staveley-Smith, Freeman, and Kilborn}}]{pisano}
\bibinfo{author}{\bibfnamefont{D.~J.} \bibnamefont{Pisano}},
  \bibinfo{author}{\bibfnamefont{D.~G.} \bibnamefont{Barnes}},
  \bibinfo{author}{\bibfnamefont{B.~K.} \bibnamefont{Gibson}},
  \bibinfo{author}{\bibfnamefont{L.}~\bibnamefont{Staveley-Smith}},
  \bibinfo{author}{\bibfnamefont{K.~C.} \bibnamefont{Freeman}},
  \bibnamefont{and} \bibinfo{author}{\bibfnamefont{V.~A.}
  \bibnamefont{Kilborn}}, \bibinfo{journal}{Astrophys. J.}
  \textbf{\bibinfo{volume}{610}}, \bibinfo{pages}{L17} (\bibinfo{year}{2004}).

\bibitem[{\citenamefont{Abadi et~al.}(2003)\citenamefont{Abadi, Navarro,
  Steinmetz, and Eke}}]{abadi}
\bibinfo{author}{\bibfnamefont{M.~G.} \bibnamefont{Abadi}},
  \bibinfo{author}{\bibfnamefont{J.~F.} \bibnamefont{Navarro}},
  \bibinfo{author}{\bibfnamefont{M.}~\bibnamefont{Steinmetz}},
  \bibnamefont{and} \bibinfo{author}{\bibfnamefont{V.~R.} \bibnamefont{Eke}},
  \bibinfo{journal}{Astrophys. J.} \textbf{\bibinfo{volume}{597}},
  \bibinfo{pages}{21} (\bibinfo{year}{2003}), \eprint{astro-ph/0212282}.

\bibitem[{\citenamefont{Kazantzidis et~al.}(2004)}]{cusps}
\bibinfo{author}{\bibfnamefont{S.}~\bibnamefont{Kazantzidis}}
  \bibnamefont{et~al.}, \bibinfo{journal}{Astrophys. J.}
  \textbf{\bibinfo{volume}{608}}, \bibinfo{pages}{663} (\bibinfo{year}{2004}),
  \eprint{astro-ph/0312194}.

\bibitem[{\citenamefont{Bennett et~al.}(2003)}]{WMAP}
\bibinfo{author}{\bibfnamefont{C.}~\bibnamefont{Bennett}} \bibnamefont{et~al.},
  \bibinfo{journal}{Astrophys. J. Suppl.} \textbf{\bibinfo{volume}{148}},
  \bibinfo{pages}{97} (\bibinfo{year}{2003}), \eprint{astro-ph/0302208}.

\bibitem[{\citenamefont{Fan et~al.}(2004)}]{Fan}
\bibinfo{author}{\bibfnamefont{X.-H.} \bibnamefont{Fan}} \bibnamefont{et~al.},
  \bibinfo{journal}{Astron. J.} \textbf{\bibinfo{volume}{128}},
  \bibinfo{pages}{515} (\bibinfo{year}{2004}), \eprint{astro-ph/0405138}.

\bibitem[{\citenamefont{de~Blok et~al.}(2001)\citenamefont{de~Blok, Zwaan,
  Dijkstra, Briggs, and Freeman}}]{deBlok:2001xz}
\bibinfo{author}{\bibfnamefont{W.~J.~G.} \bibnamefont{de~Blok}},
  \bibinfo{author}{\bibfnamefont{M.~A.} \bibnamefont{Zwaan}},
  \bibinfo{author}{\bibfnamefont{M.}~\bibnamefont{Dijkstra}},
  \bibinfo{author}{\bibfnamefont{F.~H.} \bibnamefont{Briggs}},
  \bibnamefont{and} \bibinfo{author}{\bibfnamefont{K.~C.}
  \bibnamefont{Freeman}} (\bibinfo{year}{2001}), \eprint{astro-ph/0111238}.

\bibitem[{\citenamefont{Easther et~al.}(2004)\citenamefont{Easther, Greene,
  Jackson, and Kabat}}]{Easther:2003}
\bibinfo{author}{\bibfnamefont{R.}~\bibnamefont{Easther}},
  \bibinfo{author}{\bibfnamefont{B.~R.} \bibnamefont{Greene}},
  \bibinfo{author}{\bibfnamefont{M.~G.} \bibnamefont{Jackson}},
  \bibnamefont{and} \bibinfo{author}{\bibfnamefont{D.}~\bibnamefont{Kabat}},
  \bibinfo{journal}{JCAP} \textbf{\bibinfo{volume}{0401}}, \bibinfo{pages}{006}
  (\bibinfo{year}{2004}), \eprint{hep-th/0307233}.

\bibitem[{\citenamefont{Watson and Brandenberger}(2004)}]{Watson:2003}
\bibinfo{author}{\bibfnamefont{S.}~\bibnamefont{Watson}} \bibnamefont{and}
  \bibinfo{author}{\bibfnamefont{R.}~\bibnamefont{Brandenberger}},
  \bibinfo{journal}{JHEP} \textbf{\bibinfo{volume}{03}}, \bibinfo{pages}{045}
  (\bibinfo{year}{2004}), \eprint{hep-th/0312097}.

\end{thebibliography}

\end{document}